\documentclass[useAMS,usenatbib]{mn2e}

\usepackage{float}
\usepackage{enumerate}
\usepackage{multirow}
\usepackage{array}
\usepackage{comment}
\usepackage{ifthen}
\usepackage{amssymb,amsfonts,amsmath,graphicx,multirow}
\usepackage{graphicx}
\usepackage[%
	colorlinks=true,	% false: boxed links; true: colored links
	urlcolor=blue,		% color of external links
	pdfpagelabels=true,
	hypertexnames=true,
	plainpages=false,
	naturalnames=true,
	pdftitle={Priors for Orbital Eccentricity},    % title
	pdfauthor={David M. Kipping},     % author
	linkcolor=red,          % color of internal links (change box color with linkbordercolor)
	citecolor=green,        % color of links to bibliography
        ]{hyperref}
\newboolean{color}
\setboolean{color}{true}

\newboolean{astroph}
\setboolean{astroph}{false}

\title[Priors for Orbital Eccentricity]
{Bayesian priors for the eccentricity of transiting planets}
\author[David M. Kipping]{David M. Kipping$^{1,2}$\thanks{E-mail:
dkipping@cfa.harvard.edu}\\
$^{1}$Harvard-Smithsonian Center for Astrophysics, 60, Garden Street, Cambridge, MA 02138 \\
$^{2}$Carl Sagan Fellow}
\begin{document}

\date{Accepted 2014 July 31. Received 2014 July 30; in original form 2014 March 25}

\pagerange{\pageref{firstpage}--\pageref{lastpage}} \pubyear{2014}

\maketitle

\newcommand{\multi}{{\sc MultiNest}}
\newcommand{\eccsamples}{{\tt ECCSAMPLES}}
\newcommand{\fone}{{\tt F1}}
\newcommand{\obs}{\mathrm{obs}}
\newcommand{\tru}{\mathrm{true}}
\newcommand{\KLD}{Kullback-Leibler divergence}

\label{firstpage}

\begin{abstract}

Planets on eccentric orbits have a higher geometric probability of transiting 
their host star. By application of Bayes' theorem, we reverse this logic to show 
that the eccentricity distribution of transiting planets is positively biased. 
Adopting the flexible Beta distribution as the underlying prior for 
eccentricity, we derive the marginalized transit probability as well as the 
a-priori joint probability distribution of eccentricity and argument of 
periastron, given that a planet is known to transit. These results allow to 
demonstrate that most planet occurrence rate calculations using \emph{Kepler} 
data have overestimated the prevalence of planets by $\sim10$\%. Indeed, the 
true occurrence of planets from transit surveys is fundamentally intractable 
without a prior assumption for the eccentricity distribution. Further more, we 
show that previously extracted eccentricity distributions using \emph{Kepler} 
data are positively biased. In cases where one wishes to impose an informative 
eccentricity prior, we provide a recursive algorithm to apply inverse transform 
sampling of our joint prior probability distribution. Computer code of this 
algorithm, \eccsamples, is provided to enable the community to sample directly 
from the prior
(\href{http://www.cfa.harvard.edu/~dkipping/ECCSAMPLES.html}{available here}).
\end{abstract}


\section{Introduction}
\label{sec:intro}

Orbital eccentricity is one of the most fundamental properties of a planetary 
orbit. Unlike the planets of the Solar System, exoplanet eccentricities have 
been found to exhibit broad and diverse range \citep{butler:2006,beta:2013}. 
In a Bayesian sense, our a-priori expectation of a planet's eccentricity has
dramatically shifted in the past two decades. The apparent prevalence of 
eccentric systems underscores the need for adopting data models which freely 
explore both orbital eccentricity and argument of periapsis. Inaccurate 
treatment, or a complete lack of consideration, of eccentricity leads 
to erroneous inferences of the properties of individual systems and ensemble 
populations.

The distribution of orbital eccentricities is, however, more than just a 
nuisance factor, it represents the scars of past dynamical activity in 
planetary systems \citep{rasio:1996,juric:2008,chatterjee:2008}. For this
reason, extracting the underlying eccentricity distribution of exoplanets
has become a topic of considerable recent effort (e.g. \citealt{shen:2008,
wang:2011,kane:2012}). 

Extracting an accurate eccentricity distribution demands that any observational 
biases be correctly accounted for. For transiting planets, it has been 
previously demonstrated that eccentric planets have an enhanced probability of 
eclipsing their star \citep{barnes:2007}. It therefore follows that the observed 
distribution of eccentricities from such objects is positively-biased. A 
correction of this bias requires knowledge of the probability distribution of 
eccentricity, given that we know the planet transits. Upon framing the 
probability in this manner, it is evident that the problem may be tackled via 
Bayes' theorem. Guided by this realization, we here aim to provide a suite of 
Bayesian prior probability distributions starting with the basic transit 
probability in \S\ref{sec:tran}, before tackling the priors for eccentricity and 
argument of periastron in transiting systems in \S\ref{sec:ecc}. Implications 
and suggested applications of this work are discussed in \S\ref{sec:discussion}.

\section{Prior Transit Probability}
\label{sec:tran}

\subsection{Geometric Probability}
\label{sub:geometric}

We begin by considering the prior probability that a planet transits its
host star, which is often referred to as the ``geometric'' transit probability.
This probability is crucial in the design of surveys \citep{baglin:2006,
borucki:2009,ricker:2010,rauer:2014}, predicting detection yields \citep{
beatty:2008,dzigan:2012}, as a prior when statistically validating planets 
\citep{torres:2004,fressin:2012,morton:2012} and perhaps most critically 
calculating planet occurrence rates from transit surveys \citep{hartman:2009,
youdin:2011,howard:2012,berta:2013,dong:2013,dressing:2013,fressin:2013,
petigura:2013}. Despite the importance of this probability, it is surprising
how often its effects are ignored in many of the afore mentioned tasks.

The geometric transit probability is a well-known result \citep{barnes:2007,
burke:2008,winn:2010}, although it is usually not framed in a Bayesian manner. 
In this work, we define a transiting planet to be one which satisfies the
criterion that the sky-projected impact parameter, $b$, of the transit is less 
than unity. In turn, the impact parameter is defined as the sky-projected
separation between the centre of the planet and the centre of the star in units
of the stellar radius at the instant of inferior conjunction. This instant
occurs when the true anomaly, $f$, is equal to $(\pi/2-\omega)$, where
$\omega$ is the argument of periapsis of the planet's orbit.

\begin{align}
b &= \Bigg(\frac{1-e^2}{1+e\sin\omega}\Bigg) \Bigg(\frac{ a \cos i }{R_{\star}}\Bigg),
\label{eqn:impactparam}
\end{align}

where $i$ is the orbital inclination, $a$ is the semi-major axis, $R_{\star}$
is the stellar radius and $e$ is the orbital eccentricity.
For an isotropic distribution of planetary orbits, we expect $\cos i$ to be
uniformly distributed between $0$ and $1$. Therefore, the geometric transit 
probability is computed by an integration over $\cos i$ between 
all cases where the planet transits, normalized by all cases in total,

\begin{align}
\mathrm{P}(b<1|e,\omega,(a/R_{\star})) &= \frac{ \int_{\cos(i)\,\mathrm{when}\,b=0}^{\cos(i)\,\mathrm{when}\,b=1} \mathrm{d}\cos(i) }{ \int_{\cos(i)\,\mathrm{when}\,b=0}^{\cos(i)\,\mathrm{when}\,b=\infty} \mathrm{d}\cos(i) },\nonumber\\
\qquad &= \frac{ \int_{\cos(i)=0}^{\cos(i)=\frac{1+e\sin\omega}{(a/R_{\star})(1-e^2)}} \mathrm{d}\cos(i) }{ \int_{\cos(i)=0}^{\cos(i)=1} \mathrm{d}\cos(i) },\nonumber\\
\mathrm{P}(\hat{b}|e,\omega,a_R) &= \Big(\frac{1}{a_R}\Big) \Big(\frac{1+e\sin\omega}{1-e^2}\Big),
\label{eqn:geometric}
\end{align}

where, for convenience, we have substituted $(a/R_{\star}) \to a_R$ and $(b<1) 
\to \hat{b}$.

\subsection{Independent Priors $\mathrm{P}(\omega)$ and $\mathrm{P}(e)$}

Using Bayes' theorem, one may marginalize the geometric transit probability, 
$\mathrm{P}(\hat{b}|e,\omega,a_R)$, over the conditional variables, provided 
that suitable priors may be defined. Marginalization is a powerful tool when an 
exact value for a conditional variable cannot be assigned, but the prior 
probability is known. It is worth noting that a frequentist may argue that the 
act of selecting priors removes objectivity. However, this criticism is not 
pertinent for a mature field like exoplanetary science, where informative priors 
now exist thanks to two decades of more than a thousand planet discoveries 
and/or reasonable priors are easily defined thanks to the geometrical nature of 
the problem.

As an example, one should expect that planets (i.e. all planets, not just
transiting planets) have no preferred orientation in space and so the prior 
distribution of $\omega$ may be assumed to be uniform. In this work, we
therefore adopt the following independent prior for $\omega$:

\begin{align}
\mathrm{P}(\omega) &= \frac{1}{2\pi}.
\label{eqn:wintrinsic}
\end{align}

One could adopt a uniform prior for orbital eccentricity too, representing an 
uninformative choice. However, unlike $\omega$, there are physical reasons to 
expect a non-uniform distribution in $e$, largely due to tidal circularization
\citep{trilling:2000}. Further more, as shown later, a uniform prior in $e$ is 
not amenable to marginalization. Both of these problems are overcome by using
a Beta distribution prior, as advocated in \citet{beta:2013}. The Beta 
distribution is able to reproduce a wide variety of probability distributions
despite being described compactly by just two ``shape'' parameters, $\alpha$
and $\beta$. More over, \citet{beta:2013} demonstrated that the Beta
distribution currently provides the best match to the observed distribution of
$e$ versus any previously proposed form. In this work then, we adopt the 
following independent prior for $e$:

\begin{align}
\mathrm{P}(e) &= \frac{ (1-e)^{\beta-1} e^{\alpha-1} }{ \mathrm{B}[\alpha,\beta] },
\label{eqn:eintrinsic}
\end{align}

where $\mathrm{B}[\alpha,\beta]$ is the Beta function. Note that our definition
assumes $\mathrm{P}(e)$ is the same for all planets irrespective of any other
terms such as host star properties or orbital period. On the surface, this
assumption is flawed since it is known the shorter-period planets are more
frequently circular \citep{beta:2013}, for example. In practice, this may be
easily overcome by determining $\alpha$ and $\beta$ for a sub-population and
then applying the equations derived throughout this work as normal, except with 
the understanding that all derived results only hold for the sub-population in 
question. This assumption greatly simplifies the resulting mathematics without
significant loss of applicability.

\subsection{Marginalizing $\mathrm{P}(\hat{b}|e,\omega,a_R)$}
\label{sub:tranmarg}

Equipped with independent priors for $e$ and $\omega$, one may marginalize the
geometric transit probability, $\mathrm{P}(\hat{b}|e,\omega,a_R)$ 
(Equation~\ref{eqn:geometric}) over either of these terms using Bayes' theorem.
Marginalizing over $\omega$ recovers Equation~8 of \citet{barnes:2007}:

\begin{align}
\mathrm{P}(\hat{b}|e,a_R) &= \int_{\omega=0}^{2\pi} \mathrm{P}(\hat{b}|e,a_R,\omega) \mathrm{P}(\omega) \mathrm{d}\omega,\\
\qquad&= \Big(\frac{1}{a_R}\Big) \Big(\frac{1}{1-e^2}\Big).
\label{eqn:geometric_no_w}
\end{align}

Marginalizing $\mathrm{P}(\hat{b}|e,\omega,a_R)$ over $e$ yields infinity,
if one assumes a uniform prior in $e$ over the interval $0\to1$. Adopting
the Beta distribution as the prior in $e$ yields a finite solution:

\begin{align}
\mathrm{P}(\hat{b}|\omega,a_R) &= \int_{e=0}^{1} \mathrm{P}(\hat{b}|e,a_R,\omega) \mathrm{P}(e) \mathrm{d}e,\nonumber\\
\qquad&= \Big(\frac{\Gamma[\alpha+\beta]}{a_R}\Big) \Big(\frac{ \tilde{\gamma}_{1} + \tilde{\gamma}_2 \alpha \sin\omega }{\beta-1}\Big),
\label{eqn:geometric_no_e}
\end{align}

where we have used the substitutions

\begin{align}
\tilde{\gamma}_1 &= \,_2\tilde{F}_1[1,\alpha;\alpha+\beta-1;-1],\\
\tilde{\gamma}_2 &= \,_2\tilde{F}_1[1,\alpha+1;\alpha+\beta;-1],
\end{align}

where $_2\tilde{F}_1[a,b;c;z]$ is Gauss' regularized hypergeometric function. 
One may now marginalize either Equation~\eqref{eqn:geometric_no_w} over $e$ or 
Equation~\eqref{eqn:geometric_no_e} over $\omega$ to obtain 
$\mathrm{P}(\hat{b}|a_R)$:

\begin{align}
\mathrm{P}(\hat{b}|a_R) &= \int_{e=0}^{1} \mathrm{P}(\hat{b}|e,a_R) \mathrm{P}(e|a_R) \mathrm{d}e,\nonumber\\
\qquad&= \int_{\omega=0}^{2\pi} \mathrm{P}(\hat{b}|\omega,a_R) \mathrm{P}(\omega|a_R) \mathrm{d}\omega,\nonumber\\
\qquad&= \Big(\frac{1}{a_R}\Big) \Big( \frac{\Gamma[\alpha+\beta]}{\beta-1} \Big) \tilde{\gamma}_1.
\label{eqn:geometric_no_ew}
\end{align}

Equation~\eqref{eqn:geometric_no_ew}, $\mathrm{P}(\hat{b}|a_R)$, represents
the a-priori probability that a planet transits, given its scaled semi-major
axis and accounting the effects of orbital eccentricity. This new equation
has important ramifications for expected survey yields, statistical validation
of planets and occurrence rate calculations. In the latter instance, many authors
have previously simply assumed $\mathrm{P}(\hat{b}|a_R)=a_R^{-1}$ \citep{
howard:2012,dong:2013,dressing:2013,petigura:2013} and thus not
accounted for the terms due to orbital eccentricity. For example, adopting 
$\alpha=0.867$ and $\beta=3.03$ from \citet{beta:2013} indicates that
$\mathrm{P}(\hat{b}|a_R)$ is increased by $1.13$ versus the naive circular
assumption. This in-turn means that any occurrence rate calculations assuming 
circularity will have \textit{overestimated} the true number of planets by this
factor.

\begin{figure}
\begin{center}
\includegraphics[width=8.4 cm]{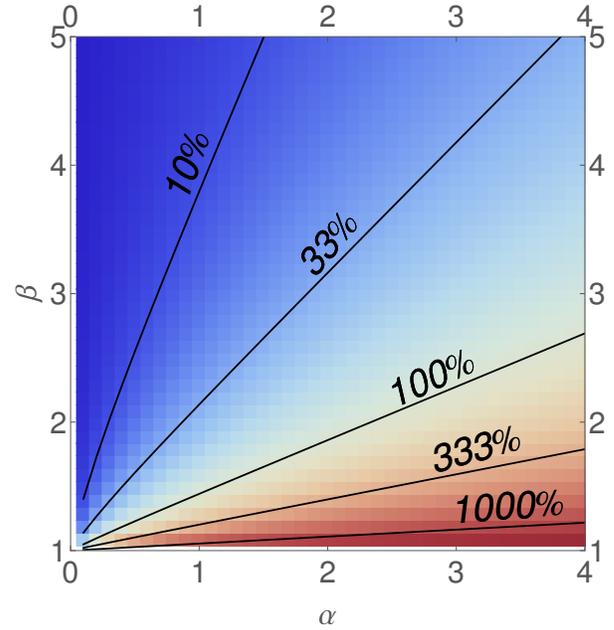}
\caption{\emph{
Enhancement of the transit probability for different shape parameters,
$\alpha$ and $\beta$, describing the underlying eccentricity distribution.
Contours mark several key enhancement levels. All loci correspond to a positive
enhancement effect.
}}
\label{fig:tranprob}
\end{center}
\end{figure}

In Figure~\ref{fig:tranprob}, we illustrate the enhancement factor in the
transit probability $\mathrm{P}(\hat{b}|a_R)$ as a function of $\alpha$ and
$\beta$, for which different values may be appropriate for different
sub-populations.

\section{Eccentricity Priors}
\label{sec:ecc}

\subsection{Orbital Eccentricity}
\label{sub:eprior}

In \S\ref{sec:tran}, we calculated the prior probability of seeing a transit,
given a specific value of the orbital eccentricity. We can reverse this 
conditional and ask the question what is the probability distribution of the 
orbital eccentricity, given that one sees transits? This question may be
answered by application of Bayes' theorem:

\begin{align}
\mathrm{P}(e|\hat{b},\omega,a_R) &= \frac{ \mathrm{P}(\hat{b}|e,\omega,a_R) \mathrm{P}(e|\omega,a_R) }{ \mathrm{P}(\hat{b}|\omega,a_R) }.
\end{align}

Note that all of the terms in the above have been already calculated. 
Specifically, $\mathrm{P}(\hat{b}|e,\omega,a_R)$ is given by 
Equation~\eqref{eqn:geometric}, $\mathrm{P}(e|\omega,a_R)$ is given by
Equation~\eqref{eqn:eintrinsic} and $\mathrm{P}(\hat{b}|\omega,a_R)$ is given by
Equation~\eqref{eqn:geometric_no_e}. Inputting these expressions reveals that 
all $a_R$ terms on the right hand side (RHS) cancel out, indicating that 
$\mathrm{P}(e|\hat{b},\omega,a_R)\to\mathrm{P}(e|\hat{b},\omega)$, where

\begin{align}
\mathrm{P}(e|\hat{b},\omega) =& \frac{ \mathrm{P}(\hat{b}|e,\omega,a_R) \mathrm{P}(e|\omega,a_R) }{ \mathrm{P}(\hat{b}|\omega,a_R) },\nonumber\\
\qquad=& \Big( \frac{ \beta-1 }{ \Gamma[\alpha+\beta] (\tilde{\gamma}_1+\tilde{\gamma}_2\alpha\sin\omega) } \Big)
\Big( \frac{ 1+e\sin\omega }{ 1-e^2 } \Big)\nonumber\\
\qquad& \Big( \frac{ (1-e)^{\beta-1} e^{\alpha-1} }{ \mathrm{B}[\alpha,\beta] } \Big).
\label{eqn:etransit}
\end{align}

\begin{figure*}
\begin{center}
\includegraphics[width=16.8 cm]{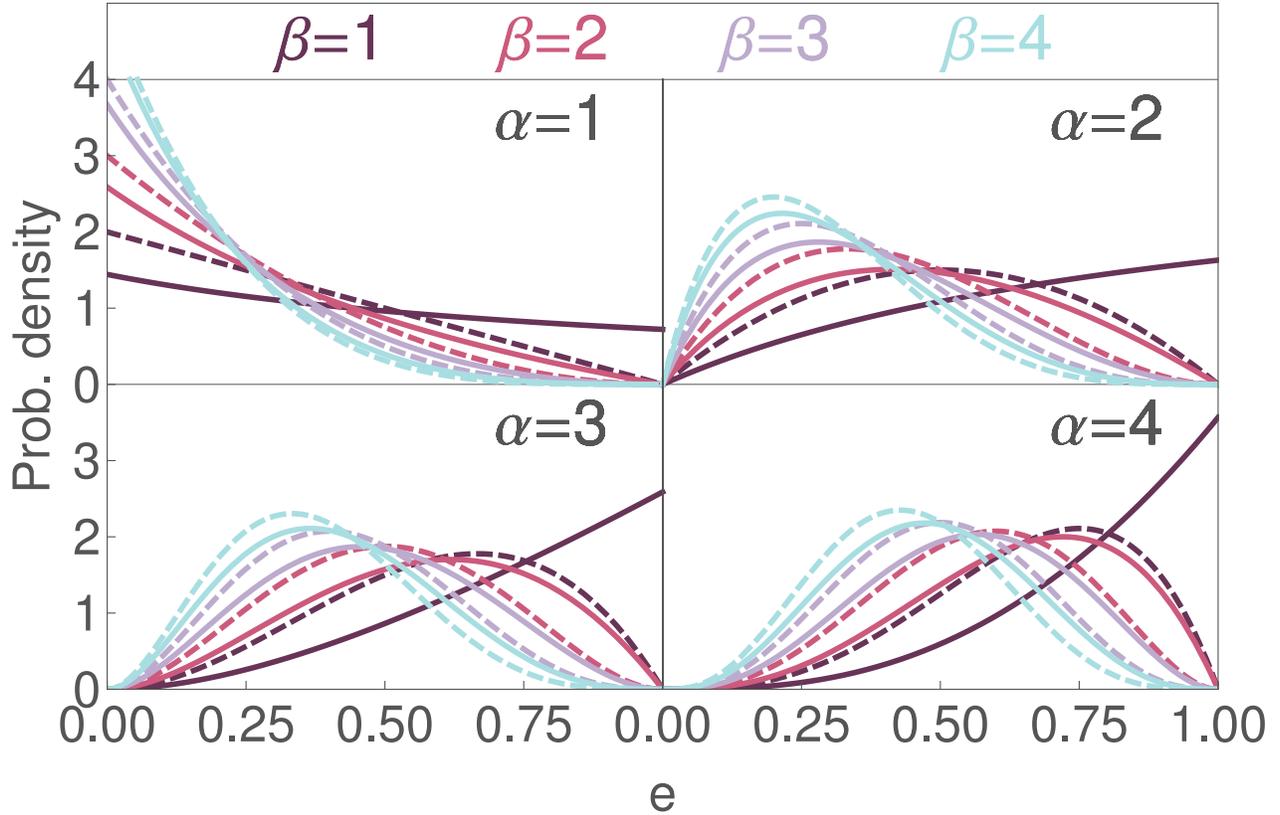}
\caption{\emph{
Illustrative comparisons of $P(e|\hat{b})$ (solid lines) versus $P(e)$ (dashed
lines). Each quadrant assumes a different $\alpha$ value, labelled in the top
right corners. Going in colour from maroon to mint, $\beta$ is varied from
$1$ to $4$ in unity steps. The largest entropy difference between the two priors
occurs for low $\beta$ and high $\alpha$.
}}
\label{fig:eprior_comp}
\end{center}
\end{figure*}

The related term $\mathrm{P}(e|\hat{b})$ may be found by a similar approach,
writing Bayes' theorem as:

\begin{align}
\mathrm{P}(e|\hat{b},a_R) &= \frac{ \mathrm{P}(\hat{b}|e,a_R) \mathrm{P}(e|a_R) }{ \mathrm{P}(\hat{b}|a_R) },\nonumber\\
\mathrm{P}(e|\hat{b}) &= \Big(\frac{ \beta-1 }{\tilde{\gamma}_1 \Gamma[\alpha+\beta] }\Big) \Big(\frac{1}{1-e^2}\Big) \Big(\frac{(1-e)^{\beta-1} e^{\alpha-1}}{\mathrm{B}[\alpha,\beta]}\Big),
\label{eqn:etransit_no_w}
\end{align}

where on the second line we drop the $a_R$ dependence which cancels out on the
RHS. We have now calculated/defined $\mathrm{P}(e|\hat{b})$ and $\mathrm{P}(e)$, 
for which the only difference is that in the former scenario we include the
conditional knowledge that the planet in question transits its host star.
Figure~\ref{fig:eprior_comp} compares these two probability density functions 
for different realizations of $\alpha$ and $\beta$, where it is visually 
apparent that $\beta\to1$ cases yield a significant entropy change.

%%% e prior comp

Consider regressing some photometric time series and/or radial velocity time
series of a planet known to transit its host star with a physical model
accounting for orbital eccentricity. Choosing an informative prior in $e$ 
naturally and quantitively includes the observer's prior knowledge about
what a typical orbital solution looks like. For an informative prior, we 
advocate using $\mathrm{P}(e|\hat{b})$ rather than $\mathrm{P}(e)$, which then
accounts for the fact a transiting planet is a-priori more likely to be 
eccentric. 

\begin{figure}
\begin{center}
\includegraphics[width=8.4 cm]{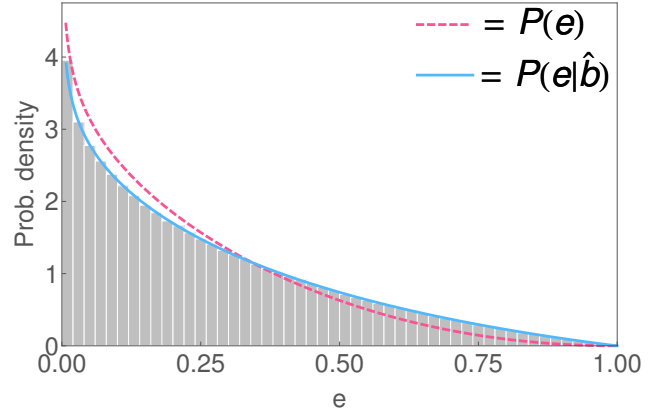}
\caption{\emph{
Histogram (gray bars) of a $10^6$ samples for $e$ computed by our \eccsamples\ 
code, which computes random variates in $\{e,\omega\}$ drawn from 
$\mathrm{P}(e,\omega|\hat{b})$, equivalent to drawing $e$ from 
$\mathrm{P}(e|\hat{b})$. The histogram provides excellent agreement to the 
closed-form PDF (solid) and clearly diverges from a prior which does not account 
for the fact the planet is transiting (dashed).
}}
\label{fig:esamples}
\end{center}
\end{figure}

In general, Monte Carlo regressions may impose priors by employing inverse
transform sampling \citep{devroye:1986}. This process allows one to generate
a uniform random variate, $x$, over the interval $[0,1]$ and transform it into
a random variate drawn from any arbitrary probability distribution. This is
achieved by finding the inverse of the cumulative density function (CDF).
For $\mathrm{P}(e|\hat{b})$, a closed-form solution for the CDF is tractable,
but the inverse is not. However, given a closed-form solution for the CDF, one 
may apply Newton's method to solve the inverse and generate samples from the 
target distribution. Accordingly, the $i^{\mathrm{th}}$ iteration of $e$ is then

\begin{align}
e_{i+1} =& e_i - \frac{1}{4} (1-e_i^2) (1-e_i)^{\beta-1} e_i^{\alpha-1} \Bigg[ 
 \Big(\frac{e_i^{\alpha+1}}{\alpha+1}\Big) \nonumber\\
\qquad& \times \Big( 2\,_2F_1[\alpha+1,2-\beta,\alpha+2;e_i] \nonumber\\
\qquad& + 3\,_2F_1[\alpha+1,1-\beta,\alpha+2;e_i] \nonumber\\
\qquad& - F_1[\alpha+1;-\beta,1;\alpha+2;e_i,-e_i] \Big) \nonumber\\
\qquad& + 4\Big( \mathrm{B}[\alpha,\beta+1;e_i]  
- x_e \mathrm{B}[\alpha,\beta-1] \tilde{\gamma}_1\Big) \Bigg],
\label{eqn:erecursive}
\end{align}

where $x_e$ is a random uniform variate, $\mathrm{B}[a,b;z]$ is the incomplete 
Beta function and $F_1[a;b,b';c;x,y]$ is Appell's hypergeometric function.
In Figure~\ref{fig:esamples}, we show a histogram of $10^6$ random variates
drawn from $\mathrm{P}(e|\hat{b})$ using this algorithm, assuming $\alpha=0.867$ 
and $\beta=3.03$. The ensemble distribution provides excellent agreement to the 
closed-form probability density function (PDF) of $\mathrm{P}(e|\hat{b})$, as 
expected.

%%% e prior

\subsection{Argument of Periastron}
\label{sub:wprior}

In \S\ref{sec:tran}, we recovered the well-known result (e.g. see 
\citealt{barnes:2007}) that the geometric probability of a planet transiting its 
host star is higher for planets with large orbital eccentricity (see 
Equations~\ref{eqn:geometric}\&\ref{eqn:geometric_no_w}). Another well-known
result for eccentric transiting planets, is that eccentric planets are most
likely to transit near periapsis ($\omega\simeq\pi/2$) than apoapsis
($\omega\simeq3\pi/2$), since the planet makes it closest approach at this
point. To our knowledge, our derived equation $\mathrm{P}(\hat{b}|\omega,a_R)$
in Equation~\eqref{eqn:geometric_no_e} is the first quantification of this 
geometric enhancement. Since periapsis transits are more likely to transit, it 
therefore follows that for planets known to transit, the prior distribution of 
$\omega$ should have a maximum at $\omega=\pi/2$ and a minimum at 
$\omega=3\pi/2$. In other words, the prior distribution of $\omega$ for 
transiting planets is manifestly non-uniform. Here, we derive this prior by 
application of Bayes' theorem; specifically

\begin{align}
\mathrm{P}(\omega|\hat{b},a_R) &= \frac{\mathrm{P}(\hat{b}|\omega,a_R) \mathrm{P}(\omega|a_R)}{\mathrm{P}(\hat{b}|a_R)}.
\end{align}

All three terms on the RHS have been defined earlier in this work, specifically
$\mathrm{P}(\hat{b}|\omega,a_R)$ is given by Equation~\eqref{eqn:geometric_no_e},
$\mathrm{P}(\omega|a_R)$ is independent of $a_R$ and thus 
$\mathrm{P}(\omega|a_R)\to\mathrm{P}(\omega)$ given by 
Equation~\eqref{eqn:wintrinsic} and finally $\mathrm{P}(\hat{b}|a_R)$ was
computed in Equation~\eqref{eqn:geometric_no_ew}. Using these expressions, the
$a_R$ dependence cancels out, indicating $\mathrm{P}(\omega|\hat{b},a_R)\to
\mathrm{P}(\omega|\hat{b})$, where

\begin{align}
\mathrm{P}(\omega|\hat{b}) &= \frac{1}{2\pi} \Big(1+\frac{\tilde{\gamma}_2 \alpha\sin\omega}{\tilde{\gamma}_1}\Big).
\label{eqn:wtransit_no_e}
\end{align}

\begin{figure}
\begin{center}
\includegraphics[width=8.4 cm]{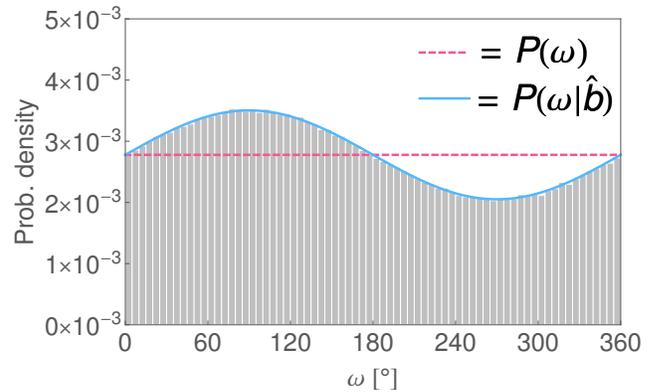}
\caption{\emph{
Histogram (gray bars) of a $10^6$ samples for $\omega$ computed by our 
\eccsamples\ code, which computes random variates in $\{e,\omega\}$ drawn from 
$\mathrm{P}(e,\omega|\hat{b})$, equivalent to drawing $\omega$ from 
$\mathrm{P}(\omega|\hat{b})$. Simulation assumes $\alpha=0.867$ and $\beta=3.03$
\citep{beta:2013}. The histogram provides excellent agreement to the 
closed-form PDF (solid) and clearly diverges from a prior which does not account 
for the fact the planet is transiting (dashed).
}}
\label{fig:wsamples}
\end{center}
\end{figure}

Inverse transform sampling may be accomplished by evaluating the inverse
CDF. Similar to the case with $\mathrm{P}(e|\hat{b})$, we were able to find a 
closed-form solution for the CDF of $\mathrm{P}(\omega|\hat{b})$, but the 
inverse yields a transcendental equation. Again, Newton's method allows for a 
simple recursive algorithm to quickly compute the inverse and thus sample 
directly from the prior $\mathrm{P}(\omega|\hat{b})$, via

\begin{align}
\omega_{i+1} =& \omega_i - 
%\Bigg(   \,_2\tilde{F}_1[1,\alpha,\alpha+\beta-1,-1] (\omega_i-2\pi x) \nonumber\\
%\qquad& + \,_2\tilde{F}_1[1,\alpha+1,\alpha+\beta,-1] \alpha (1-\cos\omega_i) \Bigg) \nonumber\\
%\qquad& \times \Bigg( \,_2\tilde{F}_1[1,\alpha,\alpha+\beta-1,-1] \nonumber\\ 
%\qquad& + \,_2\tilde{F}_1[1,\alpha+1,\alpha+\beta,-1] \alpha \sin\omega_i \Bigg)^{-1}.
\frac{\tilde{\gamma}_1 (\omega_i-2\pi x_{\omega}) + \alpha \tilde{\gamma}_2 (1-\cos\omega_i)}
{ \tilde{\gamma}_1 + \alpha \tilde{\gamma}_2 \sin\omega_i },
\label{eqn:wrecursive}
\end{align}

where $x_{\omega}$ is a random uniform variate. In Figure~\ref{fig:wsamples}, we 
plot a histogram of $10^6$ samples drawn from $\mathrm{P}(\omega|\hat{b})$, 
assuming $\alpha=0.867$ and $\beta=3.03$. The ensemble distribution provides 
excellent agreement to the closed-form PDF of $\mathrm{P}(\omega|\hat{b})$, as 
expected.

%%% w prior

Finally, it is worth noting that one can construct a prior for $\omega$ in the
case where $e$ is also known, via Bayes' theorem:

\begin{align}
\mathrm{P}(\omega|\hat{b},e,a_R) &= \frac{ \mathrm{P}(\hat{b}|e,\omega,a_R) \mathrm{P}(\omega|e,a_R) }{ \mathrm{P}(\hat{b}|e,a_R) }.
\end{align}

The terms on the RHS have been calculated/defined earlier in this work.
Specifically, $\mathrm{P}(\hat{b}|e,\omega,a_R)$ is given by 
Equation~\eqref{eqn:geometric}, $\mathrm{P}(\omega|e,a_R)$ is independent of
$a_R$ and $e$ and thus $\mathrm{P}(\omega|e,a_R)\to\mathrm{P}(\omega)$ given in
Equation~\eqref{eqn:wintrinsic} and finally $\mathrm{P}(\hat{b}|e,a_R)$ is given
by Equation~\eqref{eqn:geometric_no_w}. Using these results, we find the $a_R$ 
dependency cancels out, indicating that $\mathrm{P}(\omega|\hat{b},e,a_R)\to
\mathrm{P}(\omega|\hat{b},e)$, where

\begin{align}
\mathrm{P}(\omega|\hat{b},e) &= \frac{1+e\sin\omega}{2\pi}.
\label{eqn:wtransit}
\end{align}

\subsection{Joint Prior Probability}
\label{sub:joint}

The priors derived thus far are for the univariate case of $\mathrm{P}(e|...)$
and $\mathrm{P}(\omega|...)$. However, it is often useful to have the joint
probability density function i.e. $\mathrm{P}(e,\omega|...)$. Consider first
the simple scenario of a planet where it is unknown whether it transits or
not. Here, one may define such a prior using Bayes' theorem in two equivalent
ways:

\begin{align}
\mathrm{P}(e,\omega) &= \mathrm{P}(\omega|e) \mathrm{P}(e) \nonumber\\
\qquad&= \mathrm{P}(e|\omega) \mathrm{P}(\omega).
\end{align}

Since $\mathrm{P}(\omega|e)$ is assumed to be independent of $e$ 
(Equation~\ref{eqn:wintrinsic}) and $\mathrm{P}(e|\omega)$
is assumed to be independent of $\omega$ (Equation~\ref{eqn:eintrinsic}), then
these forms are equivalent to

\begin{align}
\mathrm{P}(e,\omega) &= \Big(\frac{1}{2\pi}\Big) \Big(\frac{(1-e)^{\beta-1}e^{\alpha-1}}{\mathrm{B}[\alpha,\beta]}\Big).
\label{eqn:ewintrinsic}
\end{align}

For transiting planets, adding in the conditional probability of $\hat{b}$
yields two equivalent forms

\begin{align}
\mathrm{P}(e,\omega|\hat{b}) &= \mathrm{P}(\omega|e,\hat{b}) \mathrm{P}(e|\hat{b}) \nonumber\\
\qquad&= \mathrm{P}(e|\omega,\hat{b}) \mathrm{P}(\omega|\hat{b}),
\end{align}

which evaluate to the same solution of

\begin{align}
\mathrm{P}(e,\omega|\hat{b}) &= \Big(\frac{\beta-1}{2\pi \tilde{\gamma}_1 \Gamma[\alpha+\beta]}\Big) \Big(\frac{1+e\sin\omega}{1-e^2}\Big) \Big(\frac{(1-e)^{\beta-1} e^{\alpha-1}}{\mathrm{B}[\alpha,\beta]}\Big).
\label{eqn:ewprior}
\end{align}

\begin{figure}
\begin{center}
\includegraphics[width=8.4 cm]{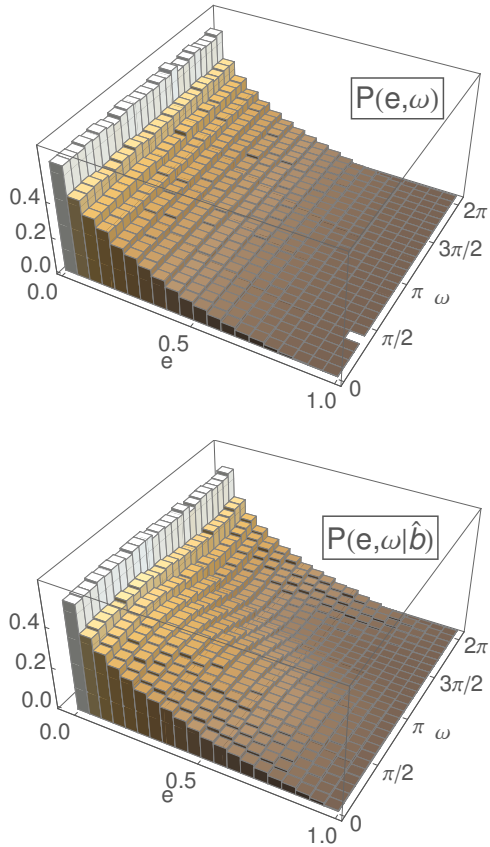}
\caption{\emph{
Three-dimensional histograms of $10^6$ samples of $\{e,\omega\}$ computed by 
\eccsamples, which reproduce $\mathrm{P}(e,\omega)$ (top) and 
$\mathrm{P}(e,\omega|\hat{b})$ (bottom). Simulations assume $\alpha=0.867$ and 
$\beta=3.03$ \citep{beta:2013}. The conditional dependence between $e$ and
$\omega$ is evident in the lower panel.
}}
\label{fig:ewhisto}
\end{center}
\end{figure}

It is important to note that the joint probability distribution is not equal to
simply multiplying the two independent priors, as was the case for planets
not known whether they transit. Specifically, whilst it is true that
$\mathrm{P}(e,\omega) = \mathrm{P}(e)\mathrm{P}(\omega)$, for transiting
planets we have $\mathrm{P}(e,\omega|\hat{b}) \neq 
\mathrm{P}(e|\hat{b})\mathrm{P}(\omega|\hat{b})$. The joint probability
distribution therefore includes the conditional dependence between $e$ and
$\omega$ (this is illustrated in Figure~\ref{fig:ewhisto}).

When comparing an observed distribution to a model distribution, it is often
useful to consider the CDF rather than PDF, in order to eliminate the effects of
subjectively choosing a bin size \citep{beta:2013}. To this end, we here provide 
the closed-form solution of the joint CDF $\mathrm{C}(e,\omega|\hat{b})$, given 
by

\begin{align}
\mathrm{C}(e,\omega|\hat{b}) &= \int_{e'=0}^{e} \int_{\omega'=0}^{\omega} \mathrm{P}(e',\omega'|\hat{b})\,\mathrm{d}e'\,\mathrm{d}\omega',\nonumber\\
\qquad&= \frac{ \gamma_3 \omega (\alpha+1) + \gamma_4 e \alpha (1-\cos\omega) }{ 2\pi e^{-\alpha}\alpha (\alpha+1) \Gamma[\alpha] \Gamma[\beta-1] },
\label{eqn:ewpriorCDF}
\end{align}

where we define

\begin{align}
\gamma_3 &= F_1(  \alpha;2-\beta,1;1+\alpha;e,-e),\\
\gamma_4 &= F_1(1+\alpha;2-\beta,1;2+\alpha;e,-e).
\end{align}

Sampling from a multivariate joint probability probability is more
elaborate than the univariate case. We first require expanding the CDF of the 
joint probability distribution using the product rule:

\begin{align}
%\mathrm{C}_{e,\omega|\hat{b}}(e',\omega') &= \mathrm{C}_{e|\hat{b}}(e') \mathrm{C}_{\omega|e,\hat{b}}(\omega'|e').
\mathrm{C}(e,\omega|\hat{b}) &= \mathrm{C}(e|\hat{b}) \mathrm{C}(\omega|e,\hat{b}).
\end{align}

The relation above indicates that one may first draw an $e$ sample, $e'$, using
the inverse of $\mathrm{C}(e|\hat{b})$ and then using this value as a 
conditional when inverting $\mathrm{C}(\omega|e=e',\hat{b})$ to obtain an 
$\omega$ sample, $\omega'$.

\begin{align}
%e &= \mathrm{C}_{e|\hat{b}}^{-1}(u_1),\\
%\omega &= \mathrm{C}_{\omega|e,\hat{b}}^{-1}(u_2).
e' &= \mathrm{C}(e|\hat{b})^{-1}[x_e],\\
\omega' &= \mathrm{C}(\omega|e=e',\hat{b})^{-1}[x_{\omega}],
\end{align}

where $x_e$ and $x_{\omega}$ are uniform random variates and $e'$ and $\omega'$
are random variates drawn from $\mathrm{P}(e,\omega|\hat{b})$. The inverse of
$\mathrm{C}(e|\hat{b})^{-1}[x_e]$ is computed using the recursive relation
presented in Equation~\eqref{eqn:erecursive}. The inverse of 
$\mathrm{C}(\omega|e,\hat{b})^{-1}$ has not yet been found and is distinct
from the inverse of $\mathrm{C}(\omega|\hat{b})^{-1}$, for which we presented
a recursive relation earlier in Equation~\eqref{eqn:wrecursive}. As before, the 
inverse requires solving a transcendental equation and again we find that 
Newton's method provides an efficient solution, using the iterations

\begin{align}
\omega_{i+1} &= \omega_i - \frac{\omega_i - 2\pi x_{\omega} + e' (1-\cos\omega_i)}{1+e'\sin\omega_i}.
\label{eqn:wrecursive2}
\end{align}

We provide Fortran code, \eccsamples\ 
(\href{https://www.cfa.harvard.edu/~dkipping/ECCSAMPLES.html}{download link}),
to to draw joint $\{e,\omega\}$ samples from $\mathrm{P}(e,\omega|\hat{b})$
using the method described above. \eccsamples\ computes $F_1$ using the \fone\ 
algorithm by \citet{f1:2004} and requires $\sim500$\,$\mu$s per sample for 1\% 
accuracy on a hyperthreaded Intel Core i7-3720QM processor. Various special
functions and dependencies required in \eccsamples\ come from 
\citet{majumder:1973}, \citet{cran:1977}, \citet{macleod:1989}, 
\citet{zhang:1996} and \citet{forrey:1997}. An example bivariate histogram of 
$10^6$ samples generated by this code is shown in Figure~\ref{fig:ewhisto}.

\section{Discussion}
\label{sec:discussion}

\subsection{Implications for Occurrence Rate Estimates}
\label{sub:implications}

It has been previously established that eccentric planets have an enhanced 
probability of transiting their host star \citep{barnes:2007}. Specifically,
one finds $\mathrm{P}(\hat{b}|e,a_R)=a_R^{-1}(1-e^2)^{-1}$; a result which we 
recover in Equation~\eqref{eqn:geometric_no_w}. However, in practice,
the thousands of transiting planet candidates found by the \emph{Kepler Mission} 
\citep{borucki:2009,batalha:2013} do not have individually measured 
eccentricities. For this reason, calculations using the ensemble population, 
such as planet occurrence rate estimates, require marginalizing out the 
eccentricity term.

The marginalized transit probability, $\mathrm{P}(\hat{b}|a_R)$, has been 
previously discussed in \citet{burke:2008}. In order to perform this 
marginalization, an assumption for the prior $\mathrm{P}(e)$ is required. Recall 
that in this work, we use the Beta distribution for $\mathrm{P}(e)$, due its 
compact yet flexible form and since it currently provides the best description 
of the observed eccentricity distribution \citep{beta:2013}. Without the 
hundreds of well-measured eccentricities available by the time of 
\citet{beta:2013}, \citet{burke:2008} used a piece-wise distribution of the form

\begin{equation}
\mathrm{P}(e)=\left\{ \begin{array}{ll}
 \frac{2}{e_{\rm crit}+e_{\rm max}}               & 0 \le e \le e_{\rm crit} \\
 \frac{2}{e_{\rm crit}+e_{\rm max}}\frac{(e-e_{\rm max})}{(e_{\rm crit}-e_{\rm max})} & e_{\rm crit} < e \le e_{\rm max} \\
    0                                          & e_{\rm max} < e < 1  \end{array}
       \right. .\label{eqn:burkeprior}
\end{equation}

Using this prior with $e_{\mathrm{crit}}=0.25$ and $e_{\mathrm{max}}=0.92$, 
\citet{burke:2008} estimate that the marginalized transit probability is 
enhanced by $\sim25$\% relative to an assumption of circularity. In this work,
we find that the transit probability is enhanced via 
Equation~\eqref{eqn:geometric_no_ew}. Drawing samples of $\alpha$ and $\beta$
from the posterior distributions derived in \citet{beta:2013}, we estimate
an enhancement factor of $13.08_{-0.71}^{+0.80}$\% for all planets in the
\citet{beta:2013} sample and $8.70_{-0.90}^{+1.08}$\% for planets in the
``short-period'' sample ($<382.3$\,d). In both cases, we favour a much lower
overall enhancement factor than \citet{burke:2008}.

In conclusion, eccentricity enhances the transit probability by $\sim10$\%,
with the exact value depending upon the sub-population in question. One 
important implication of this result is with respect to planet occurrence rate
calculations from transit surveys. In order to compute the occurrence rate,
one must account for the fact that only planets with the correct geometry
transit i.e. many more planets exist than are actually observed. We note that
many recent studies of the planet occurrence rate from the \emph{Kepler Mission}
have implicitly assumed that all planets are on circular orbits by assuming a
transit probability $\mathrm{P}(\hat{b}|a_R)=a_R^{-1}$ \citep{howard:2012,
dong:2013,dressing:2013,petigura:2013}. This effectively ignores the intrinsic
bias of the transit technique towards eccentric planets. It can therefore be 
seen that all of these afore mentioned occurrence rate calculations have 
\emph{overestimated} the true rate by $\sim10$\%.

As an example, \citet{dressing:2013} estimate $0.90_{-0.03}^{+0.04}$ planets per 
star for M-dwarfs with periods below 50\,d. Using the ``short-period'' posterior
distributions for $\alpha$ and $\beta$ computed by \citet{beta:2013}, we
estimate that this occurrence rate would be modified to $0.83_{-0.03}^{+0.04}$ 
planets per star, which is $>2$\,$\sigma$ decrease. In practice, the $\alpha$
and $\beta$ values should be derived from a sample as closely matching the
transit survey sample as possible. Here, the ``short-period'' sample of
\citet{beta:2013} spans periods up to $382.3$\,d, all spectral types and
all detectable masses. In contrast, the \citet{dressing:2013} sample only 
extends up to periods of 50\,d, stars cooler than 4000\,K and probes down to
smaller planets than that detectable by current radial velocity measurements.
Querying the Exoplanet Orbit Database \citep{wright:2011} at the time of 
writing, there are only 12 planets with radial velocity orbits for stars cooler 
than 4000\,K and periods below 50\,d, which is insufficient to extract a robust 
eccentricity distribution. This highlights the importance of measuring the 
eccentricity distributions in these poorly sampled parameter regions.

\subsection{Implications for Extracting the Eccentricity Distribution}
\label{sub:edistrib}

In \S\ref{sec:ecc}, we derived a suite of probability distributions regarding 
the eccentricity of a planet, given that we know it transits. Simply put, 
eccentric planets are more likely to transit and thus the eccentricity 
distribution of transiting planets is positively biased. To our knowledge, this 
is the first formal demonstration of this point. We also stress that this 
positive bias is distinct from another positive bias in eccentricity caused by 
the boundary condition at $e=0$ \citep{lucy:1971}. However, it is worth noting 
that this latter bias is easily avoided by using nested sampling based regression 
techniques \citep{skilling:2004}, rather than Markov chain Monte Carlo 
strategies.

The fact that eccentric planets are more likely to transit means that extracting
the eccentricity distribution of transiting planets requires a correction for 
this positive bias. Specifically, one should describe the observed distribution
of transiting planet eccentricities with $\mathrm{P}(e|\hat{b})$ rather than 
$\mathrm{P}(e)$. Recent efforts to extract the eccentricity distribution of
\emph{Kepler's} planetary candidates, such as \citet{kane:2012}, have not
accounted for this bias and so have extracted a positively skewed 
distribution.

\subsection{Using \eccsamples\ to Impose Informative Priors}
\label{sub:eccsamples}

Two of the probability distributions derived in this work may be of use to
wider community, for the purpose of serving as informative priors.
Firstly, in Equation~\eqref{eqn:etransit_no_w}, we
provide $\mathrm{P}(e|\hat{b})$, which is the a-priori probability 
distribution of a planet's eccentricity, given that it transits its host star.
Secondly, in Equation~\eqref{eqn:ewprior}, we provide 
$\mathrm{P}(e,\omega|\hat{b})$, which is the a-priori joint probability
distribution of a planet's eccentricity and argument of periastron, given that
it transits its host star.

These probability distributions may be used as priors when fitting any kind
of data of transiting planets, for which the eccentricity is a dependent 
variable. This is particularly useful in cases where the data poorly constrains
$e$ and $\omega$, yet one wishes to fully propagate the intrinsic uncertainty
in these parameters into the regression process. One example would be fitting 
low signal-to-noise radial velocity data of transiting systems, where the 
planetary mass and eccentricity exhibit strong correlations \citep{cumming:2004,
ford:2005}. Another example would be fitting transit light curves of 
any signal-to-noise level, since transit data contains very little information on 
the eccentricity \citep{kipping:2008}.

In order to facilitate the easy adoption of such priors, we provide Fortran 
code, \eccsamples, available 
\href{http://www.cfa.harvard.edu/~dkipping/ECCSAMPLES.html}{via this hyperlink}.
\eccsamples\ allows one to transform a uniform variate, $\{x_e,x_{\omega}\}$, 
into a variate drawn from $\mathrm{P}(e,\omega|\hat{b})$ i.e. $\{e,\omega\}$. In 
a typical Monte Carlo regression, one would explore parameter space as usual 
using the uniformly distributed parameters $\{x_e,x_{\omega}\}$. At each 
realization, the likelihood (or $\chi^2$) calculation is made with a trial 
$\{e,\omega\}$, which comes from transforming the current trial 
$\{x_e,x_{\omega}\}$ into $\{e,\omega\}$ using \eccsamples. Therefore, the call 
to \eccsamples\ would simply be inserted just before the likelihood call. 

\eccsamples\ can draw samples from $\mathrm{P}(e,\omega|\hat{b})$ and
$\mathrm{P}(e|\hat{b})$, as well as $\mathrm{P}(e,\omega)$ and $\mathrm{P}(e)$
for cases where it is unknown whether the planet transits. In all cases, a
fundamental assumption is that $\mathrm{P}(e)$ is described by a Beta 
distribution with shape parameters $\alpha$ and $\beta$ \citep{beta:2013}.
One may even choose to allow $\alpha$ and $\beta$ to vary, which is to say
one implements hyper-priors. \eccsamples\ therefore provides an flexible tool
for imposing eccentricity priors.

\section*{Acknowledgments}

This work was performed [in part] under contract with the California Institute 
of Technology (Caltech)/Jet Propulsion Laboratory (JPL) funded by NASA through 
the Sagan Fellowship Program executed by the NASA Exoplanet Science Institute.
This research has made use of the Exoplanet Orbit Database and the 
Exoplanet Data Explorer at exoplanets.org.

%%%%%%%%%%%%%%%%%%%%%%%%%%%%%%%%%%%%%%%%%%%%%%%%%%%%%%%%%%%%%%%%%%%%%%%%%%%%%%%%
%%%%%%%%%%%%%%%%%%%%%%%%%%%%%%%%%%%%%%%%%%%%%%%%%%%%%%%%%%%%%%%%%%%%%%%%%%%%%%%%
%%%%%%%%%%%%%%%%%%%%%%%%%%%%%%%%%%%%%%%%%%%%%%%%%%%%%%%%%%%%%%%%%%%%%%%%%%%%%%%%
%%%%%%%%%%%%%%%%%%%%%%%%%%%%%%%%%%%%%%%%%%%%%%%%%%%%%%%%%%%%%%%%%%%%%%%%%%%%%%%%
%%%%%%%%%%%%%%%%%%%%%%%%%%%%%%%%%%%%%%%%%%%%%%%%%%%%%%%%%%%%%%%%%%%%%%%%%%%%%%%%
%%%%%%%%%%%%%%%%%%%%%%%%%%%%%%%%%%%%%%%%%%%%%%%%%%%%%%%%%%%%%%%%%%%%%%%%%%%%%%%%
%%%%%%%%%%%%%%%%%%%%%%%%%%%%%%%%%%%%%%%%%%%%%%%%%%%%%%%%%%%%%%%%%%%%%%%%%%%%%%%%
%%%%%%%%%%%%%%%%%%%%%%%%%%%%%%%%%%%%%%%%%%%%%%%%%%%%%%%%%%%%%%%%%%%%%%%%%%%%%%%%
%%%%%%%%%%%%%%%%%%%%%%%%%%%%%%%%%%%%%%%%%%%%%%%%%%%%%%%%%%%%%%%%%%%%%%%%%%%%%%%%

%\clearpage
%\appendix

%\bsp

\label{lastpage}

\end{document}